\documentclass[twocolumn,showpacs,superscriptaddress,prl]{revtex4}
\usepackage[latin9]{inputenc}
\usepackage{color}
\usepackage{verbatim}
\usepackage{amsmath}
\usepackage{amssymb}
\usepackage{graphicx}
\usepackage{esint}
\usepackage{hyperref}

\makeatletter
\@ifundefined{textcolor}{}
{%
 \definecolor{BLACK}{gray}{0}
 \definecolor{WHITE}{gray}{1}
 \definecolor{RED}{rgb}{1,0,0}
 \definecolor{GREEN}{rgb}{0,1,0}
 \definecolor{BLUE}{rgb}{0,0,1}
 \definecolor{CYAN}{cmyk}{1,0,0,0}
 \definecolor{MAGENTA}{cmyk}{0,1,0,0}
 \definecolor{YELLOW}{cmyk}{0,0,1,0}
}

\usepackage{mathrsfs}

\draft
\makeatother

\begin{document}

\title{Realization of 2D Spin-orbit Interaction and Exotic Topological Orders in Cold Atoms}

\author{Xiong-Jun Liu}
\affiliation{Department of Physics, Hong Kong University of Science and Technology, Clear Water Bay, Hong Kong, China}
\affiliation{Institute for Advanced Study, Hong Kong University of Science and Technology, Clear Water Bay, Hong Kong, China}
\author{K. T. Law}
\affiliation{Department of Physics, Hong Kong University of Science and Technology, Clear Water Bay, Hong Kong, China}
\author{T. K. Ng}
\affiliation{Department of Physics, Hong Kong University of Science and Technology, Clear Water Bay, Hong Kong, China}

\begin{abstract}
Majorana zero bound mode exists in the vortex core of a chiral $p+ip$ superconductor or superfluid, which can be driven from an $s$-wave pairing state by two-dimensional (2D) spin-orbit (SO) coupling. We propose here a novel scheme based on realistic cold atom platforms to generate 2D SO interactions in a blue-detuned square optical lattice, and predict both the quantum anomalous Hall effect and chiral topological superfluid phase in the experimentally accessible parameter regimes. This work may open a new direction with experimental feasibility to observe non-Abelian topological orders in cold atom systems.
\end{abstract}
\pacs{37.10.Jk, 71.10.Pm, 67.85.Lm, 03.65.Vf}
\maketitle

\indent

{\it Introduction.}$-$
The search for non-Abelian Majorana fermions has been a focus of both theoretical and experimental studies in condensed matter physics,
driven by both the pursuit of exotic fundamental physics and the applications in
fault-tolerant topological quantum computation \cite{p-wave,TQC1,TQC2}.
Majorana zero modes (MZMs) are predicted to exist in the vortex core of a two-dimensional (2D) intrinsic $\left(p+ip\right)$-wave
superconductor (SC) and at the ends of a 1D $p$-wave
SC \cite{p-wave}. Recent studies show that 1D and 2D effective $p$-wave superconductivity can be obtained through heterostructures formed by conventional $s$-wave SCs and topological insulators \cite{TI} or spin-orbit (SO) coupled semiconductors with Zeeman splitting \cite{2D,1D}, leading to MZMs in the case of odd number of subbands crossing the Fermi level.
Following the theoretical prediction, signatures of Majorana end states have been observed in the semiconductor nanowire/$s$-wave superconductor heterostructures \cite{MFexperiment} through tunneling transport measurements \cite{DTC}. However, demonstrating the non-Abelian statistics of MZMs in braiding operation \cite{TQC1}, as an unambiguous verification, is a far more demanding task and not yet available in solid-state experiments.

On the other hand, the recent great advancement in realizing synthetic SO coupling in cold atoms \cite{Liu,Lin,Chapman,Wang,MIT,Pan} opens intriguing new avenues to probe topological orders \cite{Liu1,Wu,Goldman,Chen,1DTI,Ruostekoski,Chuanwei1,Sato,Sau1,Xiaji,Vyasanakere,BCS-BEC1,BCS-BEC2,Yun} with clean platforms in a fully controllable fashion.
So far the experimentally realized SO interaction \cite{Lin,Chapman,Wang,MIT,Pan}  is a 1D SO term with equal Rashba and Dresselhaus amplitudes through a two-photon Raman process as theoretically proposed in Ref. \cite{Liu}. Detailed investigations show that only 1D SO term can be realized in an atom gas in the continuum with an internal $\Lambda$-type configuration as considered in experiments \cite{Liu,Ye,Victor}. With such 1D SO interaction it is unfortunately not optimistic to reach the topological superfluid (SF) phase from an $s$-wave pairing state in cold atoms. The reason is because on one hand in a quasi-1D system no long-range $s$-wave order can be obtained; on the other hand, the proximity effect, as used in solid state experiments \cite{MFexperiment}, is not realistic for cold atoms. Therefore, to find out a truly experimental scheme to observe MZMs with higher dimensional SO interactions is a foremost outstanding goal in the current field of cold atoms.

In this letter, we propose a novel scheme to observe both the quantum anomalous Hall effect (QAHE) and chiral topological SF phase with (pseudo)spin-$1/2$ cold fermions trapped in a blue-detuned square optical lattice and coupled to two periodic Raman fields. The advantage of our model in the realization is that the square optical lattice and Raman fields, used to induce 2D SO interactions, are generated through minimum number of laser fields. Basing our model on real cold atom systems and conventional technologies, we show that the predicted exotic topological orders can be studied with currently available experimental platforms.

\begin{figure}[t]
\includegraphics[width=1.0\columnwidth]{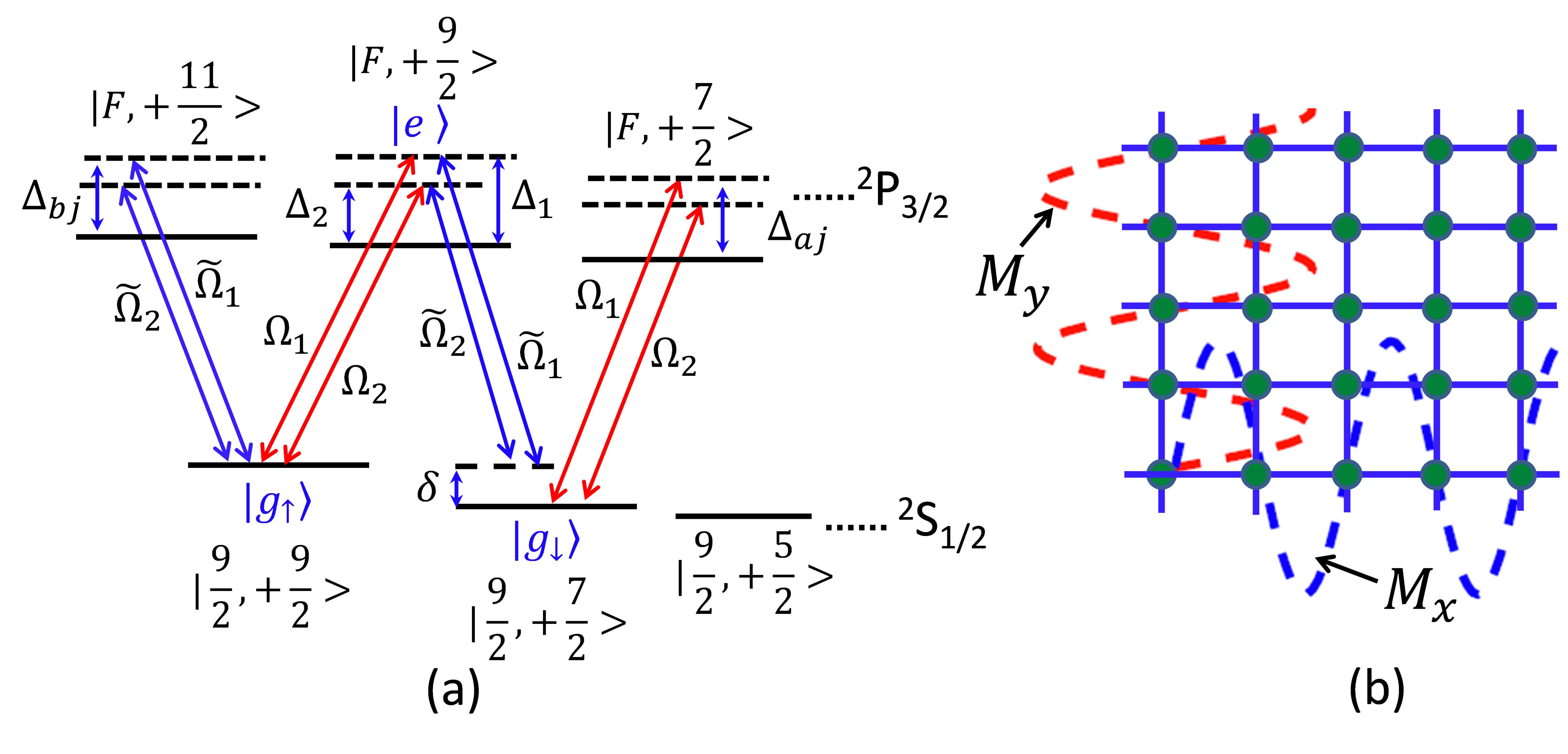}\caption{(Color online) (a) A realistic optical-dipole transition diagram in cold fermions ($^{40}$K) coupled to two pairs of laser beams $\Omega_{1,2}$ and $\tilde\Omega_{1,2}$ under blue-detuning condition. (b) With this configuration a square lattice potential and two periodic Raman fields $M_{x,y}$ are simultaneously generated.}
\label{lattice}
\end{figure}

{\it Model.}$-$We start with the quasi-2D cold fermions trapped in a conventional square optical lattice, with their internal degree of freedom (atomic spins) experiencing periodic Raman fields $M_{x,y}$ induced by two-photon processes, as illustrated in Fig.~\ref{lattice}. The dynamics of fermions are governed by the following Hamiltonian
\begin{eqnarray}\label{eqn:H1}
H&=&\frac{p_x^2}{2m}+\frac{p_y^2}{2m}+V(\bold r)+m_z(|g_\uparrow\rangle\langle g_\uparrow|-|g_\downarrow\rangle\langle g_\downarrow|)\nonumber\\
&&-\bigr\{\bigr[M_x(x)+iM_y(y)\bigr]|g_\uparrow\rangle\langle g_\downarrow|+{\rm H.c.}\bigr\},
\end{eqnarray}
where $M_x=M_0\sin(k_0x)$ and $M_y=M_0\sin(k_0y)$, and the optical dipole potential $V(\bold r)=-V_0\bigr[\cos^2(k_0x)+\cos^2(k_0y)\bigr]$ forms the square lattice. Below we describe how to obtain $H$ with the realistic cold atom platform (details of the realization can be found in the supplementary material~\cite{SI}). The diagram for light-atom couplings is shown in Fig.~\ref{lattice}(a), where four blue-detuned lasers are applied to induce the couplings, with two standing-wave lasers $\Omega_{1,2}$ linearly polarized along $z$ axis and propagating in the $x$-$y$ plane, and two $\sigma_+$-polarized plane-wave lasers $\tilde\Omega_{1,2}$ propagating along $z$ direction. According to the selection rule, the standing-wave and plan-wave lasers can induce the transitions from ground states $|g_{\uparrow,\downarrow}\rangle$ to excited ones depicted by red and blue lines in Fig.~\ref{lattice}(a), respectively. For the present purpose, we set $\Omega_1=\Omega_0\sin(k_0x)$ and $\Omega_2=\Omega_0\sin(k_0y)$, while $\tilde\Omega_1=i\tilde\Omega_2=\Omega_0$ to be constants~\cite{SI}. In the parameter regime that $\Delta_{1,2}\gg|\Delta_1-\Delta_2|\gg|\Omega_0^2/\Delta_{1,2}|$, the population of the excited state $|e\rangle$ is negligible \cite{Liu}, and the effective model of ground states is governed by two main effects induced by these transitions. First, the atom-light couplings contribute to diagonal potentials for $|g_{\uparrow,\downarrow}\rangle$ given by $V_\uparrow=\sum_{j=1,2}\hbar(|{\Omega}_j|^2/\Delta_j+|\tilde{\Omega}_{j}|^2/\Delta_{b_j})$ and $V_\downarrow=\sum_j\hbar(|{\tilde\Omega}_j|^2/\Delta_j+|{\Omega}_j|^2/\Delta_{a_j})$, which create the square lattice. Second, note that the couplings between $|g_{\uparrow,\downarrow}\rangle$ and $|e\rangle$ consist of two $\Lambda$-type configurations, through which two independent Raman fields are generated. In particular, $\Omega_1,\tilde\Omega_1$ generate one Raman field $M_x=\hbar|\Omega_0|^2\sin(k_0x)/\Delta_1$, and $\Omega_2,\tilde\Omega_2$ generate another one given by $iM_y=i\hbar|\Omega_0|^2\sin(k_0y)/\Delta_2$. The Raman processes can be precisely tuned in experiment to have a small two-photon off-resonance $\delta$ ($|\delta|\ll|\Omega_0^2/\Delta_{1,2}|$) which gives rise to the constant Zeeman term with $m_z=\hbar\delta/2$ in $H$ \cite{1DTI}. We note that while the applied lasers $\Omega_{1,2}$ and $\tilde{\Omega}_{1,2}$ can also couple to other ground states [e.g. $|9/2,+5/2\rangle$ in Fig.~\ref{lattice}(a)], these couplings cannot lead to additional Raman transitions between $|g_{\uparrow,\downarrow}\rangle$ and other ground levels due to large two-photon detunings which are much greater than Raman fields $|\Omega_0^2/\Delta_{1,2}|$ \cite{Wang,MIT,SI}, and thus are neglected. In the realistic experiment (for $^{40}$K atoms) the magnitudes of detunings $\Delta_{a,b,j}$ (can be over $1.0$THz) are much larger than their differences (in the order of $10\sim100$MHz) \cite{Wang}. Therefore in the formulas of $V_\sigma$ and $M_{x,y}$ we can take that $\Delta_1=\Delta_2=\Delta_{a_{1,2}}=\Delta_{b_{1,2}}=\Delta$, which is followed by $V_\uparrow=V_\downarrow=-(\hbar|\Omega_0|^2/\Delta)\bigr[\cos^2(k_0x)+\cos^2(k_0y)\bigr]+{const.}$ and $M_0=\hbar|\Omega_0|^2/\Delta$. Neglecting the constant terms yields the effective Hamiltonian~\eqref{eqn:H1}.

Before proceeding further, we provide two important remarks on the realization. First, since both the square lattice and the periodic spatial profile of the Raman fields are determined by the same standing wave lasers $\Omega_{1,2}$, the phase fluctuations in the standing waves, characterized by $\Omega_1=\Omega_0\sin(k_0x+\phi_1^{\rm fluc})$ and $\Omega_2=\Omega_0\sin(k_0y+\phi_2^{\rm fluc})$, only lead to global shift of the lattice and Raman fields as illustrated in Fig.~\ref{lattice}(b). The relative spatial profile of $M_{x,y}$ and $V(\bold r)$ is always automatically fixed and therefore the effective Hamiltonian~\eqref{eqn:H1} is unchanged. This greatly simplifies the experimental setup in the realization. Second, the blue detuning is essential for the realization. If using red-detuned lasers, i.e. $\Delta_{a,b,j}<0$, one has then $V_{\uparrow,\downarrow}=-(|\Omega_0^2/\Delta|)\bigr[\sin^2(k_0x)+\sin^2(k_0y)\bigr]$, which shifts $1/2$ lattice site relative to the Raman fields. The red-detuned laser couplings cannot lead to 2D SO interaction or nontrivial topological orders as studied below.

{\it Quantum anomalous Hall effect.}$-$The tight-binding model of $H$ can be derived straightforwardly. We take that fermions occupy the lowest $s$-orbitals $\phi_{s\sigma}$ ($\sigma=\uparrow,\downarrow$), and consider only the nearest-neighbor hopping terms. The tight-binding Hamiltonian is given by
$H_{\rm TI}=-t_s\sum_{<\vec{i},\vec{j}>,\sigma}\hat c_{\vec{i}\sigma}^{\dag}\hat
c_{\vec{j}\sigma}+\sum_{\vec{i}}m_z(\hat n_{\vec{i}\uparrow}-\hat n_{\vec{i}\downarrow})+\bigr[\sum_{<\vec{i},\vec{j}>}t_{\rm so}^{\vec{i}\vec{j}}\hat c_{\vec{i}\uparrow}^\dag\hat c_{\vec{j}\downarrow}+{\rm H.c.}]$, where $t_s$ denotes spin-conserved hopping, the 2D lattice-site index $\vec i=(i_x,i_y)$, and $\hat n_{\vec{i}\sigma}=\hat c_{\vec{i}\sigma}^\dag\hat c_{\vec{i}\sigma}$. From the even-parity of the $s$-orbitals and the periodic profile of Raman fields shown in Fig.~\ref{lattice}(b), it can be directly verified that the spin-flip hopping terms due to the Raman fields satisfy $t_{\rm so}^{j_x,j_x\pm1}=\pm(-1)^{j_x}t_{\rm so}^{(0)}$ and  $t_{\rm so}^{j_y,j_y\pm1}=\pm i(-1)^{j_y}t_{\rm so}^{(0)}$, where $t_{\rm so}^{(0)}=M_0\int d^2\vec r\phi_{s}(x,y)\sin(k_0x)\phi_{s}(x-a,y)$ with $a$ the lattice constant~\cite{SI}.
Redefining the spin-down operator $\hat c_{\vec{j}\downarrow}\rightarrow e^{i\pi\vec r_j/a}\hat c_{\vec{j}\downarrow}$, we recast the Hamiltonian into
\begin{eqnarray}\label{eqn:tightbinding2}
H_{\rm TI}&=&-t_s\sum_{<\bar{i},\vec{j}>}(\hat c_{\vec{i}\uparrow}^{\dag}\hat
c_{\vec{j}\uparrow}-\hat c_{\vec{i}\downarrow}^{\dag}\hat
c_{\vec{j}\downarrow})+\sum_{\vec{i}}m_z(\hat n_{\vec{i}\uparrow}-\hat n_{\vec{i}\downarrow})+\nonumber\\
&&+\bigr[\sum_{j_x}t_{\rm so}^{(0)}(\hat c_{j_x\uparrow}^\dag\hat c_{j_x+1\downarrow}-\hat c_{j_x\uparrow}^\dag\hat c_{j_x-1\downarrow})+{\rm H.c.}\bigr]+\nonumber\\
&&+\bigr[\sum_{j_y}it_{\rm so}^{(0)}(\hat c_{j_y\uparrow}^\dag\hat c_{j_y+1\downarrow}-\hat c_{j_y\uparrow}^\dag\hat c_{j_y-1\downarrow})+{\rm H.c.}\bigr].
\end{eqnarray}
It is convenient to rewrite $H_{\rm TI}$ in the $k$-space and $H_{\rm TI}=-\sum_{\bold k,\sigma\sigma'}\hat c_{\bold k,\sigma}^{\dag}[d_z(\bold k)\sigma_z+d_x(\bold k)\sigma_x+d_y(\bold k)\sigma_y]_{\sigma,\sigma'}\hat
c_{\bold k,\sigma'}$, with $d_x=2t_{\rm so}^{(0)}\sin(k_ya)$, $d_y=2t_{\rm so}^{(0)}\sin(k_xa)$ and $d_z=-m_z+2t_s\cos(k_xa)+2t_s\cos(k_ya)$. A simple analysis shows that the bulk system is gapped when $|m_z|\neq4t_s,0$. By calculating the first Chern number $C^{\rm AH}_1=(4\pi)^{-1}\int d^2\bold k\bold n\cdot\partial_{k_x}\bold n\times\partial_{k_y}\bold n$, with $\bold n=(d_x,d_y,d_z)/|\vec d(\bold k)|$, one can verify that the above Hamiltonian describes a topological insulator for $0<|m_z|<4t_s, t_{\rm so}^{(0)}\neq0$ and otherwise a trivial insulator. This result can be read by analyzing the band structure around the four independent Dirac points at $\bold k_c=(0,0), (0,\pi),(\pi,0),$ and $(\pi,\pi)$, where the system are described by massive Dirac equations. First, it is straightforward to know that the system is a trivial insulator when $m_z\gg4t_s$. Upon reducing $m_z$ to $m_z=4t_s$ the band gap closes at $\bold k_c=(0,0)$ and reopens when $0<m_z<4t_s$, with the effective mass at this Dirac point changing sign and leading to the change by $1$ in the Chern number, and therefore the system becomes topological with $C_1^{\rm AH}=1$. In the same way, one can determine that the phase is topological for $-4t_s<m_z<0$. Note that masses of both Dirac Hamiltonians around $\bold k_c=(0,\pi)$ and $(\pi,0)$ change sign when varying $m_z<0$ to $m_z>0$, implying that $C_1^{\rm AH}(m_z=0_+)-C_1^{\rm AH}(m_z=0_-)=2$ and therefore $C_1^{\rm AH}=-1$ for $-4t_s<m_z<0$. We then conclude that the Hamiltonian \eqref{eqn:tightbinding2} realizes a QAHE with the quantized Hall conductance
\begin{eqnarray}\label{eqn:conductivity}
\sigma_{xy}^{\rm AH}= \left\{ \begin{array}{ll}
         \mbox{sgn}(m_z)\frac{1}{\hbar}, \ \ \ \mbox{for} \ 0<|m_z|<4t_s, t_{\rm so}^{(0)}\neq0,\\
        0, \ \ \ \ \ \ \ \ \ \ \ \ \ \ \mbox{otherwise}.\\
        \end{array} \right.
\end{eqnarray}

It is noteworthy that while the QAHE was previously studied in spinless cold fermion models \cite{Wu}, it requires complicated lattice configurations and staggered gauge potentials which are hard to be achieved in the experiments. Here we have proposed a novel scheme to observe QAHE in spin-$1/2$ fermion system which is truly accessible with current experimental platforms.

{\it Chiral topological superfluid phase.}$-$We next study the topological chiral superfluid phase by considering an $s$-wave interaction which can be well controlled with Feshbach resonance in cold atoms \cite{RMPcoldatom}. In optical lattice, this interaction is described by the attractive Fermi Hubbard model $H_{\rm int}=-\sum_{\vec i}Un_{\vec i\uparrow}n_{\vec i\downarrow}$, where the regularized 2D effective interaction $U=\hbar^2a_s\sqrt{8\pi}/(ml_zl_{\rm 2d}^2)$ with $l_z=(h/m\omega_z)^{1/2}$, $l_{\rm 2d}=(h/m\omega_{\rm 2d})^{1/2}$, and the s-wave scattering length $a_s>0$. The parameters $\omega_{2d}$ and $\omega_z$ denote the 2D square lattice trapping frequency and a tight trapping frequency along $z$ axis, respectively. We first study the superfluid phase with the self-consistent mean field approach by introducing the s-wave superfluid order parameter by $\Delta_s=(U/N_0)\sum_{\bold k}\langle c_{\bold k\uparrow}c_{-\bold k\downarrow}\rangle$ with $N_0$ the number of lattice sites. Then the total Hamiltonian in the Nambu basis $\psi_{\bold k}=(c_{\bold k\uparrow},c_{\bold k\downarrow},c^\dag_{-\bold k\downarrow},-c^\dag_{-\bold k\uparrow})$ can be written as $H_{\rm BdG}=\sum_{\bold k}\psi_{\bold k}^\dag{\cal H}_{\rm BdG}(\bold k)\psi_{\bold k}$, where
\begin{eqnarray}\label{eqn:TSF1}
{\cal H}_{\rm BdG}&=&d_x\sigma_x\otimes\tau_z+d_y\sigma_y\otimes\tau_z+d_z\sigma_z\otimes I-\mu I\otimes\tau_z+\nonumber\\
&&+(\Delta_s I\otimes\tau_++{\rm H.c.}).
\end{eqnarray}
Here $\mu$ is the chemical potential, $\tau_{x,y,z}$ are Pauli matrices acting on the Nambu space, and $\tau_\pm=(\tau_x\pm i\tau_y)/2$. The $s$-wave order parameter is solved by the gap equation
\begin{eqnarray}\label{eqn:gap}
\frac{N_0}{U}=\frac{1}{4}\sum_{\bold k,\alpha=\pm}\frac{1}{E_{\bold k}^{\alpha}}\bigr[1+\alpha\frac{d_z^2(\bold k)}{w_{\bold k}}\bigr]\tanh(\frac{1}{2}\beta E^\alpha_{\bold k}),
\end{eqnarray}
where the energy spectra $E^{\pm}_{\bold k}=(\Gamma^2+|\vec d_{\bold k}|^2\pm2w_{\bold k})^{1/2}$ by diagonalizing ${\cal H}_{\rm BdG}$, with $w_{\bold k}=[\mu^2(d_x^2+d_y^2)+d_z^2\Gamma^2]^{1/2}$ and $\Gamma^2=\mu^2+|\Delta_s|^2$, and $\beta=1/(k_BT)$.

To identify an effective $p+ip$ chiral superfluid phase from the Hamiltonian ${\cal H}_{\rm BdG}$, we analyze the special situation with $m_z=4t_s$. In this case the low energy spectrum is captured by the Hamiltonian (up to linear order of the momentum) ${\cal H}_{\rm BdG}=2t_{\rm so}^{(0)}(k_x\sigma_y+k_y\sigma_x)\tau_z-\mu\tau_z+(\Delta_s\tau_++{\rm H.c.})$. It is interesting that this low-energy model is equivalent to the surface Hamiltonian of a 3D topological insulator in proximity to an s-wave SC \cite{TI}, which renders the 2D chiral $p+ip$ topological SC phase. We therefore expect that in our model the chiral Majorana edge modes and non-Abelian MZM in the vortex core can be obtained \cite{Tewari}.
Similar as the case in QAHE, the SF phase diagram can also be determined by analyzing the properties of the bulk gap.
By examining $E^{-}_{\bold k}$ with $m_z>0$ we find that the bulk gap closes at $\bold k=(0,0)$ and $\bold k=(\pi,\pi)$ for $\mu^2+|\Delta_s|^2=(m_z-4t_s)^2$ and $(m_z+4t_s)^2$, respectively. Furthermore, at the points $\bold k=(0,\pi)$ and $\bold k=(\pi,0)$ the bulk gap closes if $\mu^2+|\Delta_s|^2=m_z^2$. These properties lead to a rich phase diagram with $t_{\rm so}^{(0)}\neq0$ as presented below. From the low-energy Hamiltonian ${\cal H}_{\rm BdG}$ we know that the superfluid phase with $|\Delta_s|\gtrsim0$, $m_z=4t_s$, and $\mu=0$ is topological. This implies that if $|m_z|>2t_s$, the SF phase is topologically nontrivial with Chern number $C_1^{\rm SF}=+1$ for $(|m_z|-4t_s)^2<\mu^2+|\Delta_s|^2<m_z^2$, and $C_1^{\rm SF}=-1$ for $m_z^2<\mu^2+|\Delta_s|^2<(|m_z|+4t_s)^2$. The change by $2$ in the Chern number is originated from the fact that tuning $\mu^2+|\Delta_s|^2$ from less than to larger than $m_z^2$ reverses the mass terms at both $\bold k=(0,\pi)$ and $\bold k=(\pi,0)$.
On the other hand, when $m_z=0$ the phase is always trivial. Actually, when $m_z=0$ the low-energy Hamiltonian ${\cal H}_{\rm BdG}$ for the SF is captured by two Dirac cones around $\bold k=(0,\pi)$ and $\bold k=(\pi,0)$ with opposite chiralities and having $s$-wave pairing. Such two Dirac cones resemble $p+ip$ and $p-ip$ superfluids, respectively, which couple to each other and cancel out, yielding a trivial phase. With this result we know that if $|m_z|<2t_s$, the SF phase is topologically trivial for $m_z^2<\mu^2+|\Delta_s|^2<(|m_z|-4t_s)^2$, while it is nontrivial with $C_1^{\rm SF}=-1$ when $(|m_z|-4t_s)^2<\mu^2+|\Delta_s|^2<|m_z|+4t_s)^2$. In cold atoms the parameters such as $m_z, t_s$ and $\mu$ can be precisely adjusted in the experiment to tune the system into topological chiral SF phases.

\begin{figure}[t]
\includegraphics[width=1.0\columnwidth]{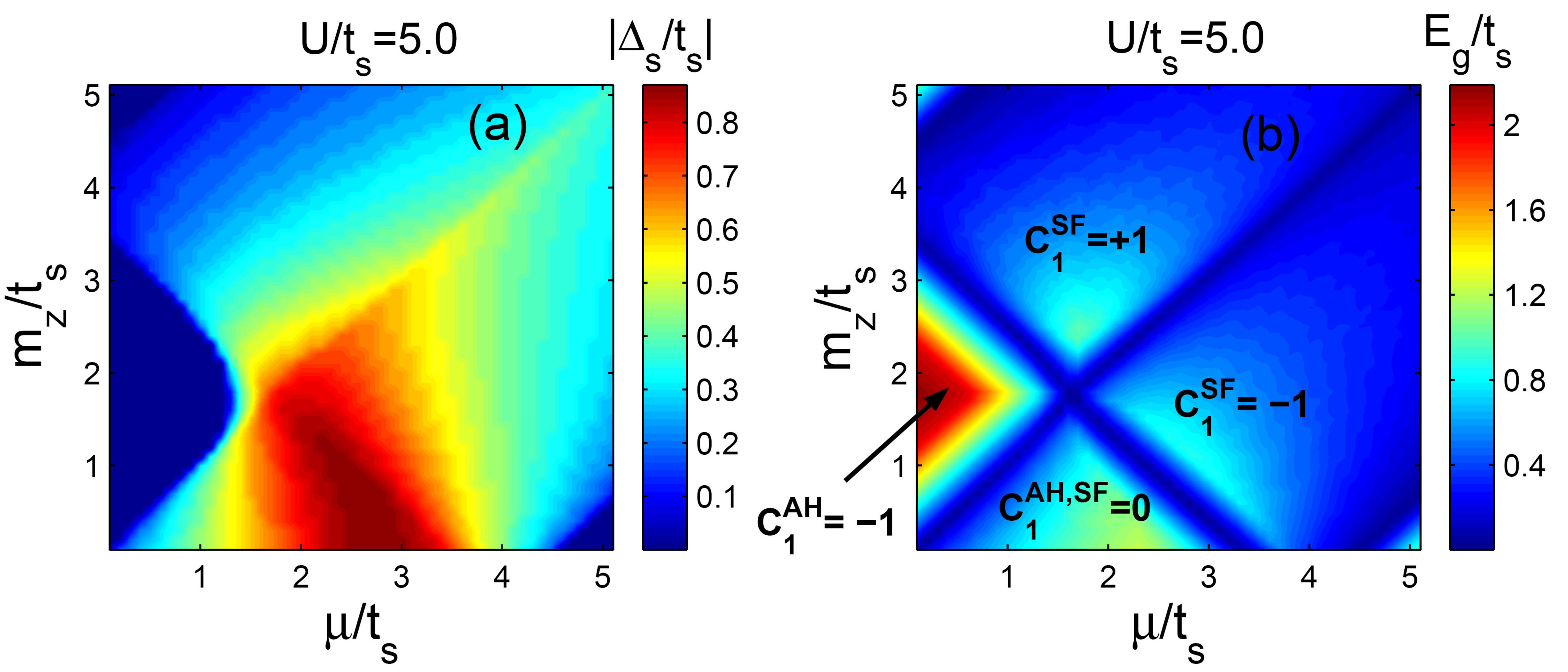}\caption{(Color online) Self-consistent calculations of the $s$-wave order $\Delta_s$ [for (a)] and the bulk gap $E_g$ [for (b)] versus $m_z$ and $\mu$ at zero temperature. Topological phase transition occurs at $E_g=0$. The hopping coefficients $t_{\rm so}^{(0)}=t_s$.}
\label{meanfield}
\end{figure}
The self-consistent solutions for $\Delta_s$ and mean-field phase diagram are shown in Fig.~\ref{meanfield} (a) and (b), respectively. It can be seen that the $s$-wave order $\Delta_s$ vanishes when the noninteracting Hamiltonian $H_{\rm TI}$ has a large insulating gap and the chemical potential $\mu$ locates deep in the band gap [left most area in (a)]. On the other hand, an appreciable $\Delta_s$ is obtained in the area with $1.5t_s<\mu<3.5t_s$ and $0<m_z<3.0$. In such parameter regime the SO terms of $d_x$ and $d_y$ dominates over $d_z$ and a relatively large density of states at Fermi level is present. In Fig.~\ref{meanfield} (b) the superfluid phases with both $C_1^{\rm SF}=+1$ and $-1$ are obtained, respectively leading to chiral and anti-chiral Majorana edge modes localized in the boundary \cite{p-wave}.

Note that in 2D superfluids no long-range order exists at finite temperature and the critical temperature by mean-field theory is often overestimated. Instead, the Berezinsky-Kosterlitz-Thouless (BKT) transition occurs at the critical temperature which is limited by entropically driven vortex and antivortex proliferation \cite{BKT}. The BKT temperature is calculated by
\begin{eqnarray}\label{eqn:BKT}
T_{\rm BKT}=\frac{\pi}{2}\rho_s(\Delta_s,T_{\rm BKT}),
\end{eqnarray}
where $\rho_s$ is the superfluid stiffness (superfluid density). In the presence of phase fluctuation, the superfluid order parameter takes the form $\Delta_s=\Delta_0e^{i\theta(\bold r)}$, where $\theta$ varies slowly in the position space. The simplest way to estimate the stiffness is to use its relation to the supercurrent density $\bold j_s=\rho_s\nabla\theta(\bold r)$. For the case that $\theta$ varies slowly with position, we can approximate that $\bold q=\nabla\theta(\bold r)$, which implies that the $s$-wave pairing occurs between two fermions with a center-of-mass momentum $\bold q$. The current can also be calculated by variation of the free energy $\bold j_s=\frac{1}{A}{\rm Tr}[e^{-\beta H_{\rm BdG}}\delta H_{\rm BdG}(\mu,\Delta_0,\nabla\theta)/\delta\bold q]/{\rm Tr}[e^{-\beta H_{\rm BdG}}]$, with A the area of the lattice.
Solving this equation and the gap equation for $\Delta_s$ together we can get the stiffness by $\rho_s=\bold j_s/\bold q$. It can be verified that these results are consistent with those through a standard functional path integral approach, where the phase fluctuation of the superfluid order is described by an effective action $S_{\rm fluc}=(1/2)\int d^2\bold r\rho_s(\nabla\theta)^2$. The BKT temperature is then given by Eq.~\eqref{eqn:BKT}. We have numerically confirmed that when $\Delta_s=0$ the supercurrent $\bold j_s$ vanishes.

\begin{figure}[ht]
\includegraphics[width=1.0\columnwidth]{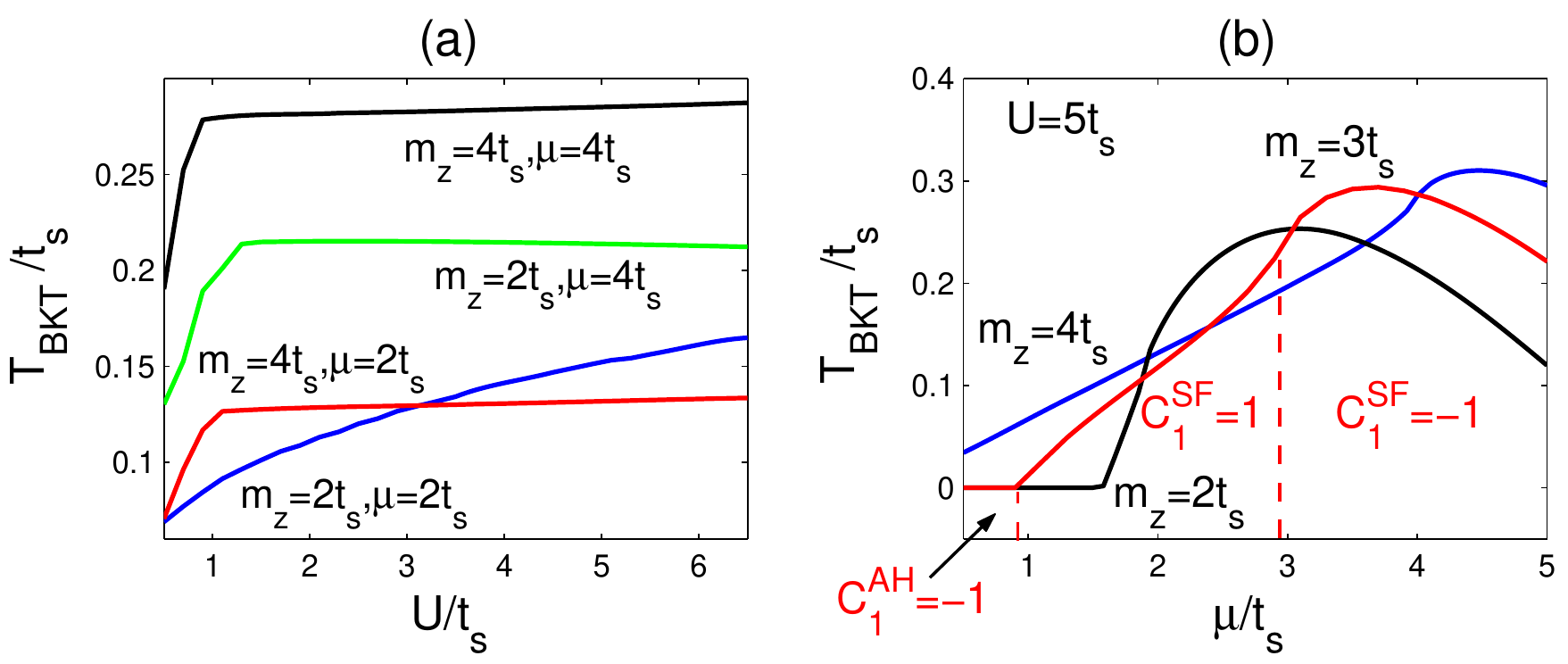}\caption{(Color online) BKT temperature as a function of $U$ with different values of $m_z$ and $\mu$ (a), and versus $\mu$ at $U=5t_s$ and with different magnitudes of $m_z$ (b). The SO coupled hopping $t_{\rm so}^{(0)}=t_s$. Topological orders of different Chern numbers are indicated for the case with $m_z=3t_s$.}
\label{BKT}
\end{figure}
The numerical results for $T_{\rm BKT}$ versus $U$ and $\mu$ are shown in Fig.~\ref{BKT}. For the case that the Fermi energy touches the bulk edge, the BKT temperature increases gradually with $U$ [blue line in Fig.~\ref{BKT}(a)], while when the Fermi energy locates deep into the bulk, the $T_{\rm BKT}$ increases sharply with $U$ and approaches a saturate magnitude in a small Hubbard interaction (red, green, and black curves). The BKT temperature versus $\mu$ with fixed $U=5t_s$ and $m_z$ is plotted in Fig.~\ref{BKT}(b). It can be found that with $m_z=3t_s$, the maximum $T_{\rm BKT}$ is around $0.3t_s$. Note that the upper value of the blue detuning is limited by the fine-structure splitting and can be taken as $2\pi\times1.7$THz for $^{40}$K atoms \cite{Wang,Tiecke}, which gives the recoil energy $E_R/\hbar\sim\hbar k_0^2/2m=2\pi\times8.5$kHz using lasers of wavelength $764$nm to form the square lattice. Taking $\Omega_0=2\pi\times0.27$GHz, we have $V_0=5E_R$, the lattice trapping frequency $\omega=2\pi\times34.9$kHz, and the hopping coefficients $t_{\rm so}^{(0)}/\hbar\sim t_s/\hbar\simeq2\pi\times0.52$kHz. With this parameter regime we find that the bulk gap for QAHE $E_g=2\pi\times2.08$kHz by setting $\delta=4t_s$, corresponding to the temperature $T\approx0.33T_F\approx100$nK for observation, and $T_{\rm BKT}$ for the topological superfluid phase with $U=5t_s,\delta=6t_s$, and $\mu=-3.5t_s$ is about $0.086T_F$.

{\it Conclusions and Discussions.}$-$We have proposed a truly experimental scheme to observe 2D SO interaction which leads to both the quantum anomalous Hall effect and chiral topological superfluid phase in spin-$1/2$ cold fermions trapped with a blue-detuned square optical lattice and coupled to two periodic Raman fields. We calculated the bulk gap of the topological states and the BKT temperature of the topological superfluid phase, and show that the predicted topological orders are reachable with the realistic parameter regimes.

The present model exhibits many essential advantages in the realization. First of all, the square optical lattice and the periodic Raman fields, used to induce 2D SO interactions, are generated through the same standing-wave lasers, which greatly simplifies the setup for experimental studies. On the other hand, for blue-detuned optical lattice the atoms are trapped in the minimums of lattice potentials which may generally minimize heating effects \cite{blue}. Note that while the topological superfluid phase with Majorana modes requires a fermion system, the topology and band structure of the insulating state can be demonstrated with cold bosons \cite{Atala}, which makes experimental observations even more straightforward. It is worthwhile to point out that 2D chiral topological orders are classified by a $Z$ invariant, and is stable when stacking multi-layers of the 2D system. This enables a study of topological superfluid in the many-layer square lattice system which may have a higher transition temperature since the fluctuation in the superfluid order can be suppressed. The feasibility of the present work will motivate future experimental studies on the 2D topological orders and the observation of non-Abelian Majorana modes with realistic cold atom platforms.


We appreciate Gyu-boong Jo, Waseem Bakr, and Lawrence Cheuk much for very helpful discussions on the experimental feasibility of this work. We also thank Wujie Huang, H. Zhai, M. Cheng, Patrick A. Lee, C. Wu, and Z.-X. Liu for insightful discussions. We acknowledge the support from HKRGC through Grant 605512 and HKUST3/CRF09.


\noindent

\onecolumngrid

\renewcommand{\thesection}{S-\arabic{section}}
\renewcommand{\theequation}{S\arabic{equation}}
\setcounter{equation}{0}  
\renewcommand{\thefigure}{S\arabic{figure}}
\setcounter{figure}{0}  

\section*{\Large\bf Supplementary Information}

In this supplementary material we provide the details of deriving the effective Hamiltonians~(1-2) in the main text. This derivative is based on the experimental data of realistic cold atoms. We shall show that both the conventional square lattice and the Raman fields are realized by applying the four laser fields $\Omega_{1,2}$ and $\tilde\Omega_{1,2}$, with the former two $\Omega_{1,2}$ being standing waves set in the $x$-$y$ plane and the latter two $\tilde\Omega_{1,2}$ the plane waves propagating along $z$ direction. In particular, the two Raman fields are induced by the four lasers ($\Omega_1,\tilde\Omega_1$ generate one Raman field, and $\Omega_2,\tilde\Omega_2$ generate another independent Raman field), while the square lattice is created by the two standing waves. Note that both the square lattice and the spatial profile of the Raman fields are determined by the same standing wave lasers $\Omega_{1,2}$. The relative configuration between the Raman fields and square lattice is automatically fixed, which can greatly simplify the experimental setup.

\section{Effective Hamiltonian}

For clarity, we describe the model step by step. We first demonstrate the conventional square lattice generated by the two standing waves $\Omega_{1,2}$, and then show that the Raman fields are induced when the two additional lasers $\tilde\Omega_{1,2}$ are applied.

\subsection{Spin-independent lattice potential}
\begin{figure}[ht]
\includegraphics[width=0.8\columnwidth]{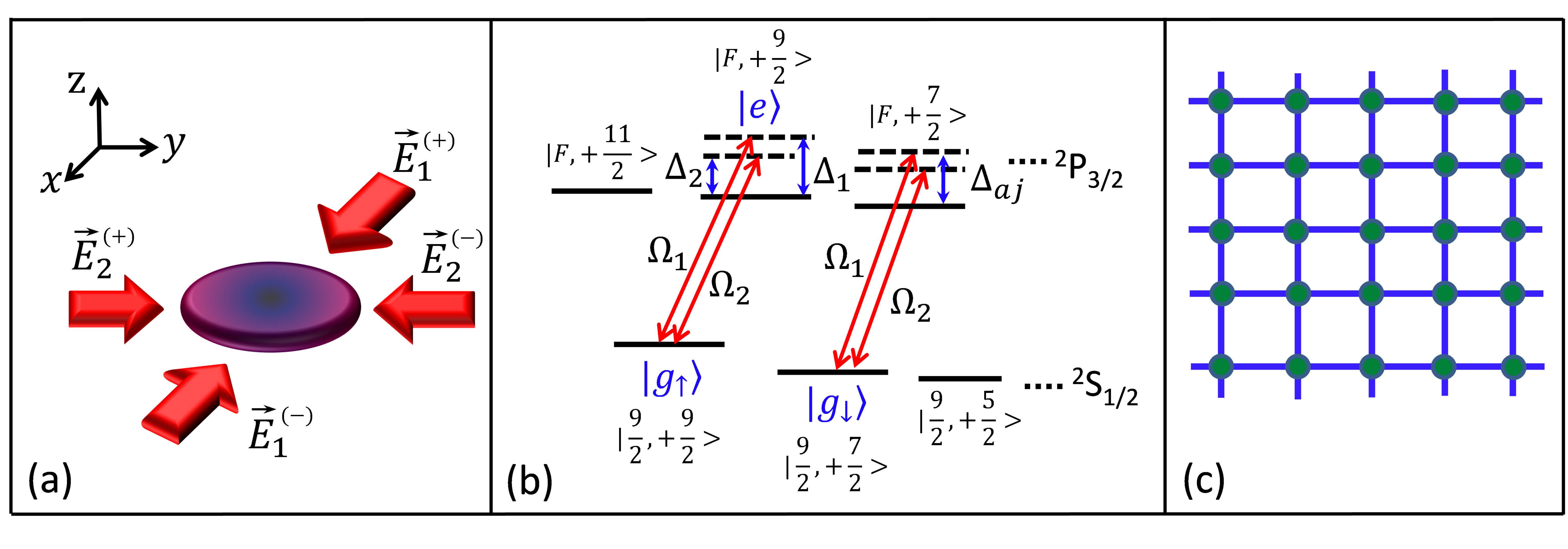} \caption{(Color online) Square optical lattice generated by $\pi$ transitions in the realistic $^{40}$K atoms. (a) Two linearly polarized standing wave lasers $\vec E_{1,2}$, with each formed by forward- and backward-propagating components, are set along $x$ and $y$ axes to generate the square lattice potential. (b) With the polarization in the $z$ direction, the lasers induce transitions from ground states $|g_{\uparrow,\downarrow}\rangle$ to excited ones with the same $m_F$. For the regime with single-photon detunings much larger than the hyperfine-structure splitting but less than the fine-structure splitting, the effective Rabi-frequencies and the lattice potential are spin-independent. (c) The configuration of the generated square optical lattice.}
\label{SImodel1}
\end{figure}
We generate the square lattice potential from the standing wave lasers, described by their electric fields $\vec E_{1,2}$, with standard procedures used in optical lattice experiments, as illustrated in Fig.~\ref{SImodel1}. Our purpose here is to show that the lattice potentials are spin-independent with the present laser configuration and in the experimental parameter regime for generating spin-orbit interaction. The standing wave lasers $\vec E_1=\vec E_1^{(+)}+\vec E_1^{(-)}=\vec E_{01}\sin(k_0x)$ and $\vec E_2=\vec E_2^{(+)}+\vec E_2^{(-)}=\vec E_{02}\sin(k_0y)$ are applied along $x$ and $y$ directions, respectively. Here $\vec E_j^{(+)}$ and $\vec E_j^{(-)}$ are the forward- and backward-propagating components of the standing wave $\vec E_j$ [Fig.~\ref{SImodel1} (a)]. The $\pi$ transitions illustrated in Fig.~\ref{SImodel1} (b), which conserve $m_F$, can be achieved by choosing linearly polarized lasers $\vec E_{1,2}$ with the polarization along $z$ direction. Similar as the standard square lattice realization \cite{SIRMPcoldatom}, a finite difference between the deunings $\delta\omega=\Delta_1-\Delta_2$ ($\sim10$MHz to $100$MHz) is used to avoid interference between the two standing waves. Such difference can also lead to independent Raman transitions when the other two plane wave lasers are applied (see next subsection). Note that $\vec E_{1,2}$ can drive the $\pi$ transitions from the ground states $|\frac{9}{2},+\frac{9}{2}\rangle$ and $|\frac{9}{2},+\frac{7}{2}\rangle$ to all possible excited levels which satisfy the selection rule. For the $D2$ lines in the $^{40}$K atoms, the state $|\frac{9}{2},+\frac{9}{2}\rangle$ ($|g_\uparrow\rangle$) can couple to two excited levels $|\frac{11}{2},+\frac{9}{2}\rangle$ and $|\frac{9}{2},+\frac{9}{2}\rangle$, while the state $|\frac{9}{2},+\frac{7}{2}\rangle$ ($|g_\downarrow\rangle$) can couple to three excited levels $|\frac{11}{2},+\frac{7}{2}\rangle$, $|\frac{9}{2},+\frac{7}{2}\rangle$, and $|\frac{7}{2},+\frac{7}{2}\rangle$ \cite{SITiecke}. The detunings $\Delta_{1,2}$ (can be in the order of $1.0$THz) are much larger than the hyperfine-structure splitting (a few tens of MHz), but should not exceed the fine-structure splitting which is the frequency difference between the $D1$ and $D2$ lines (about $2\pi\times1.7$THz). In this regime, the $D1$ transitions induced by the applied lasers will not affect our effective model (see below for further discussion). The optical dipole potentials for $|g_{\uparrow,\downarrow}\rangle$ can be obtained by summing over the contributions by all allowed transitions, and are denoted by $V_{\uparrow,\downarrow}=V^{(1)}_{\uparrow,\downarrow}(x)+V^{(2)}_{\uparrow,\downarrow}(y)$, where
\begin{eqnarray}\label{eqn:diople}
V^{(1)}_\uparrow(x)&=&\frac{|\Omega_1^{\uparrow}|^2}{\Delta_1}, \ V^{(2)}_\uparrow(y)=\frac{|\Omega_2^{\uparrow}|^2}{\Delta_2}, \\
V^{(1)}_\downarrow(x)&=&\frac{|\Omega_1^{\downarrow}|^2}{\Delta_{a1}}, \ V^{(2)}_\downarrow(y)=\frac{|\Omega_1^{\downarrow}|^2}{\Delta_{a2}}.
\end{eqnarray}
Here we have neglected the differences in the detunings for transitions to different excited hyperfine levels, since they are much less than $\Delta_{1,2}$. Actually, it is easy to know that the differences of all deunings in the above formulas can be ignored and in the following we simply take all of them to be $\Delta$ when calculating optical potentials. The effective Rabi-frequencies $\Omega^{\uparrow,\downarrow}_{1,2}$ for the lasers are defined through
\begin{eqnarray}\label{eqn:Rabi1}
|\Omega_1^{\uparrow}(x)|&=&\sqrt{\sum_F|\Omega^{(9/2)}_{1F}|^2}, \ |\Omega_2^{\uparrow}(y)|=\sqrt{\sum_F|\Omega^{(9/2)}_{2F}|^2}, \ \ F=\frac{11}{2}, \frac{9}{2}, \\
|\Omega_1^{\downarrow}(x)|&=&\sqrt{\sum_F|\Omega^{(7/2)}_{1F}|^2}, \ |\Omega_2^{\downarrow}(y)|=\sqrt{\sum_F|\Omega^{(7/2)}_{2F}|^2}, \ \ F=\frac{11}{2}, \frac{9}{2}, \frac{7}{2}.
\end{eqnarray}
where ($j=1,2$)
\begin{eqnarray}\label{eqn:Rabi}
|\Omega^{(9/2)}_{jF}|=\frac{|\mu_{9/2,F}||\vec E_j|}{\hbar}, \ \ \mu_{9/2,F}=\langle\frac{9}{2},+\frac{9}{2}|\bold r\cdot\hat\epsilon_j|F,+\frac{9}{2}\rangle,\\
|\Omega^{(7/2)}_{jF}|=\frac{|\mu_{7/2,F}||\vec E_j|}{\hbar}, \ \ \mu_{7/2,F}=\langle\frac{9}{2},+\frac{7}{2}|\bold r\cdot\hat\epsilon_j|F,+\frac{7}{2}\rangle,
\end{eqnarray}
with $\mu_{9/2,F}$ and $\mu_{7/2,F}$ the dipole matrix elements, and $\hat\epsilon_j$ denoting the polarization vectors of the lasers. Below by using the experimental data for $^{40}$K atoms we demonstrate that $\Omega_1^\uparrow(x)=\Omega_1^\downarrow(x)$ and $\Omega_2^\uparrow(y)=\Omega_2^\downarrow(y)$, in the parameter regime that $\Delta_{1,2}$ are large compared with the hyperfine-structure splitting in the $D_2$ line, but less than the fine-structure splitting. Namely, the optical lattice potentials are spin-independent, as required in our proposal.

The data of the transition strength, proportional to $|\mu_{9/2,F}|^2$ or $|\mu_{7/2,F}|^2$ for the corresponding coupling can be found in Ref.~\cite{SITiecke}. It is straightforward to know that
\begin{eqnarray}\label{eqn:diople1}
\frac{V^{(1)}_\uparrow(x)}{V^{(1)}_\downarrow(x)}=\frac{V^{(2)}_\uparrow(y)}{V^{(2)}_\downarrow(y)} =\frac{|\mu_{9/2,11/2}|^2+|\mu_{9/2,9/2}|^2}{|\mu_{7/2,11/2}|^2+|\mu_{7/2,9/2}|^2+|\mu_{7/2,7/2}|^2}.
\end{eqnarray}
Using the experimental data of the relative transition strength for $^{40}$K atoms~\cite{SITiecke}, we have from the above formula that
\begin{eqnarray}\label{eqn:diople2}
\frac{V^{(1)}_\uparrow(x)}{V^{(1)}_\downarrow(x)}=\frac{V^{(2)}_\uparrow(y)}{V^{(2)}_\downarrow(y)} =\frac{1215+3240}{2187+1960+308}=1.
\end{eqnarray}
Therefore, the lattice potentials are spin-independent and from now on we can denote that $\Omega_j=\Omega_j^\uparrow=\Omega_j^\downarrow$ with $j=1,2$ [Fig.~\ref{SImodel1} (b)]. Actually, for $\pi$ transitions in $^{40}$K atoms induced by linearly polarized lasers the optical potentials (Rabi-frequencies) are always spin-independent when the detunings are large compared with the hyperfine structure splitting. Taking that laser strengths $|\vec E_{01}|=|\vec E_{02}|$, we further obtain that
\begin{eqnarray}\label{eqn:diople3}
V(\bold r)=V^{(1)}(x)+V^{(2)}(y)=-V_0\bigr[\cos^2(k_0x)+\cos^2(k_0y)\bigr],
\end{eqnarray}
where $V_0=|\Omega_0|^2/\Delta$, with $\Omega_0$ the the amplitude of the effective Rabi-frequencies $\Omega_{1,2}$ induced by the standing wave lasers. A constant term has been omitted in the above formula.

Note that we have ignored the $D1$ transitions induced by the two standing waves. When the blue detunings $\Delta_{1,2}$ are comparable with the fine-structure splitting, the $D1$ transitions actually contribute to the lattice potential and can lead to a small spin-dependent term in the lattice potentials. However, by a similar numerical estimate based the experimental data for $^{40}$K \cite{SITiecke}, we have verified that the spin-dependent term of the lattice potentials due to the $D1$ transitions is only slightly above $2\%$ of the spin-independent term when the detunings $\Delta_{1,2}$ is less than the fine-structure splitting. Therefore the effective model in this proposal is unaffected by $D1$ transitions.

\subsection{Raman fields}

\begin{figure}[]
\includegraphics[width=0.8\columnwidth]{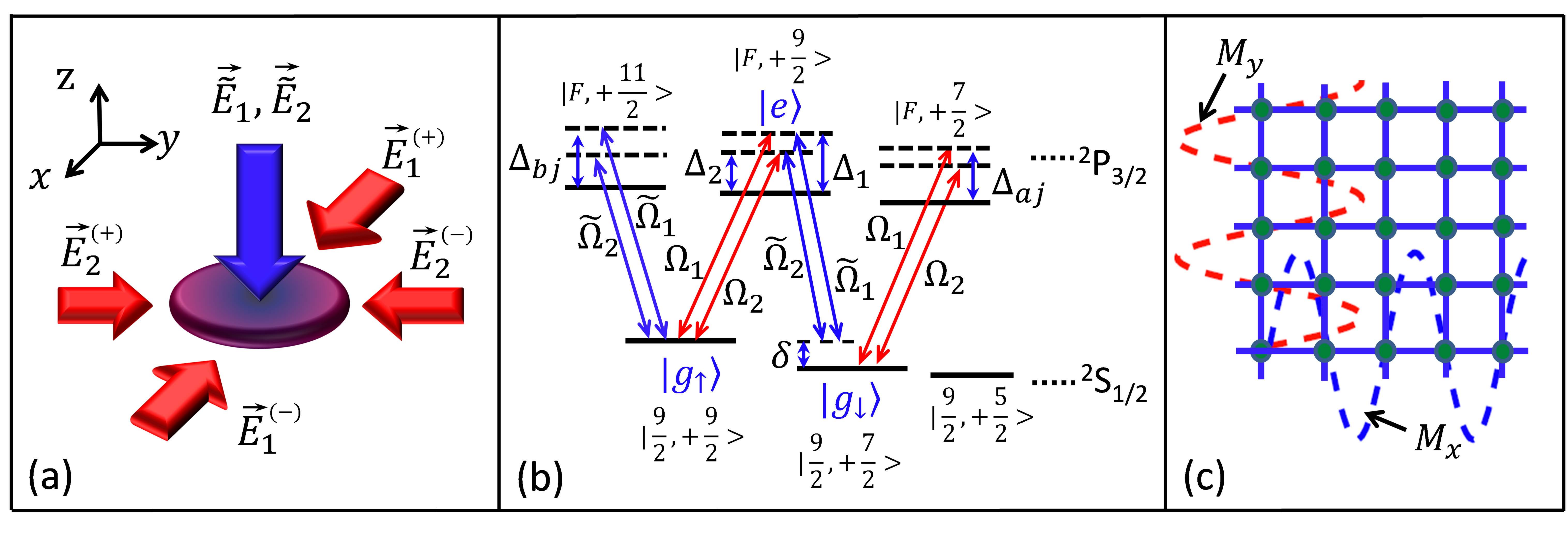} \caption{(Color online) (a) Based on the setup shown in Fig.~\ref{SImodel1}, two additional plane-wave $\sigma_+$ lasers $\vec{\tilde E}_{1,2}$ are applied along $z$ direction. (b) With the four lasers two independent Raman fields are generated through two-photon processes. (c) The periodicity of the Raman fields is double of the lattice period, and the Raman fields are antisymmetric corresponding to the center of each lattice site. This relative configuration is automatically fixed, since the spatial profiles of both the square lattice and the Raman fields are determined by the same standing wave lasers $\Omega_{1,2}$.}
\label{SImodel2}
\end{figure}
Now we study the generation of the Raman fields by adding another two plane wave lasers propagating along $z$ direction [Fig.~\ref{SImodel2} (a)] to induce the transitions shown in Fig.~\ref{SImodel2} (b). The applied lasers $\vec{\tilde E}_{1,2}$ can have $\sigma_+$ polarization to induce the illustrated transitions from state with $m_F$ to the others with $m_F+1$ (Linear polarized lasers with polarization along $x$ or $y$ direction can also induce the required couplings, and the effective model of the present proposal can be realized, although in this case additional transitions from $m_F$ to $m_F-1$ will also be induced). Similar to the previous subsection, the two lasers can drive all possible transitions satisfying the selection rule. As we consider the $D2$ line in the $^{40}$K atoms, the ground state $|\frac{9}{2},+\frac{9}{2}\rangle$ can couple to the excited level $|\frac{11}{2},+\frac{11}{2}\rangle$, while the state $|\frac{9}{2},+\frac{7}{2}\rangle$ can couple to two excited levels $|\frac{11}{2},+\frac{9}{2}\rangle$ and $|\frac{9}{2},+\frac{9}{2}\rangle$ \cite{SITiecke}. By a similar calculation, the effective Rabi-frequencies induced by the two plane waves on the ground states satisfy
\begin{eqnarray}\label{eqn:Ramandiople1}
\frac{|\tilde\Omega_1^\uparrow|^2}{|\tilde\Omega_1^\downarrow|^2}=\frac{|\tilde\Omega_1^\uparrow|^2}{|\tilde\Omega_2^\downarrow|^2} =\frac{|\tilde\mu_{9/2,11/2}|^2}{|\tilde\mu_{7/2,11/2}|^2+|\tilde\mu_{7/2,9/2}|^2}.
\end{eqnarray}
Here the dipole matrix elements are defined by $\mu_{9/2,F}=\langle\frac{9}{2},+\frac{9}{2}|\bold r\cdot\hat\epsilon_j|F,+\frac{11}{2}\rangle$ and $\mu_{7/2,F}=\langle\frac{9}{2},+\frac{7}{2}|\bold r\cdot\hat\epsilon_j|F,+\frac{9}{2}\rangle$.
Using the experimental data of the relative transition strength for $^{40}$K atoms~\cite{SITiecke}, we have from the above formula that
\begin{eqnarray}\label{eqn:Ramandiople2}
\frac{|\tilde\Omega_1^\uparrow|^2}{|\tilde\Omega_1^\downarrow|^2}=\frac{|\tilde\Omega_1^\uparrow|^2}{|\tilde\Omega_2^\downarrow|^2} =\frac{13365}{10935+1440}=1.08.
\end{eqnarray}
This implies a small spin-dependent diagonal potential induced by $\vec{\tilde E}_{1,2}$, which modifies the Zeeman term $m_z\sigma_z$ in the effective Hamiltonian, or equivalently, shifts the energies of the two ground states (Stark shift). Furthermore, as mentioned in the previous subsection, the two $\sigma_+$ lasers $\vec{\tilde E}_{1,2}$ can also drive $D1$ transition from the ground state $|\frac{9}{2},+\frac{7}{2}\rangle$ to the excited one $|\frac{9}{2},+\frac{9}{2}\rangle$ in the manifold $^2$P$_{1/2}$ (not shown in Fig.~\ref{SImodel2}) \cite{SITiecke}. Taking this contribution into account we further find for $\Delta_{1,2}\simeq2\pi\times1.7$THz that
\begin{eqnarray}\label{eqn:Ramandiople2}
\frac{|\tilde\Omega_1^\uparrow|^2}{|\tilde\Omega_1^\downarrow|^2}=\frac{|\tilde\Omega_1^\uparrow|^2}{|\tilde\Omega_2^\downarrow|^2} =\frac{13365}{10935+1440+720}=1.02.
\end{eqnarray}
Therefore the actual spin-dependent potential results in an effective Zeeman term (or energy shift) with the magnitude being about $0.01(|\tilde\Omega_{1}|^2+|\tilde\Omega_{2}|^2)/\Delta$, with $|\tilde\Omega_j|=|\tilde\Omega_j^\uparrow|\simeq|\tilde\Omega_j^\downarrow|$. It is interesting that this effect can provide a useful way to effectively adjust the constant Zeeman term $m_z$. With these results in the following we again can approximate the effective Rabi-frequencies induced by $\vec{\tilde E}_{1,2}$ to be spin-independent to study the Raman fields.

The Raman fields are generated by two-photon processes \cite{SILiu,Wang,MIT}. When $|\Delta_{1,2}|\gg|\Omega_j|,|\tilde\Omega_j|$, the population of excited states due to single-photon transitions is ignorable. Then the two-photon processes dominate and can drive the Raman couplings between two ground states $|g_\uparrow\rangle$ and $|g_\downarrow\rangle$ if the two-photon off-resonance $\delta$ is small compared with the coupling strengths given by $|\Omega_j\tilde\Omega_j|/\Delta$. Note that the difference of the single-photon detunings $\delta\omega=\Delta_1-\Delta_2$ (about $10\sim100$MHz) is much greater than $|\Omega_j\tilde\Omega_j|/\Delta$ ($\sim0.1$MHz). This leads to two independent Raman transitions, with one induced by lasers $\Omega_1,\tilde\Omega_1$, and another by $\Omega_2,\tilde\Omega_2$ [Fig.~\ref{SImodel2} (b)]. On the other hand, while from the selection rule the four lasers can couple to all ground levels in the manifold $^{2}$S$_{1/2}$ ($F=9/2$), these couplings cannot drive additional Raman transitions between $|g_{\uparrow,\downarrow}\rangle$ and other ground states (e.g. $|9/2,+5/2\rangle$ in the ground manifold). This is because in the experiment of alkali atoms, the degeneracy of the ground states is non-uniformly lifted by the external magnetic field. For $^{40}$K atoms, with a magnetic field of strength $31$G and tuning the Raman transition between $|g_\uparrow\rangle$ and $|g_\downarrow\rangle$ to be nearly resonant, the Raman couplings between $|g_{\uparrow,\downarrow}\rangle$ and other ground levels will be completely suppressed by large two-photon detunings ($\sim10$MHz) \cite{SIWang}. As a result, the Raman couplings only occur between $|g_\uparrow\rangle$ and $|g_\downarrow\rangle$, with the Raman fields given by
\begin{eqnarray}\label{eqn:Raman1}
M_x(x)=\frac{\Omega_1\tilde\Omega_1^*}{\Delta}, \ M_y(y)=-i\frac{\Omega_2\tilde\Omega_2^*}{\Delta},
\end{eqnarray}
In the above form we have neglected the contributions of Raman couplings through $D_1$ lines, which slightly reduce $M_{x,y}$ under the condition that $\Delta_{1,2}$ are less than the fine-structure splitting.
To induce the 2D spin-orbit coupling, we require a $\pi/2$-phase shift between the two plane waves $\tilde\Omega_1$ and $\tilde\Omega_2$, which can be locked with mature technologies in the experiment. Practically, the two plane waves $\tilde\Omega_{1,2}$ can originate from the a single laser using beam splitter and an acousto-optic modulator (AOM) to precisely control their frequency difference (similar for $\Omega_{1,2}$). The AOM can lock their relative phase at the same time. By setting that $\tilde\Omega_1=i\tilde\Omega_2=\Omega_0$, we have
\begin{eqnarray}\label{eqn:Raman1}
M_x(x)=\frac{|\Omega_0|^2}{\Delta}\sin(k_0x), \ M_y(y)=\frac{|\Omega_0|^2}{\Delta}\sin(k_0y).
\end{eqnarray}
With the generated lattice potential and the Raman fields, we finally reach the effective Hamiltonian for the present model by
\begin{eqnarray}\label{eqn:H1}
H&=&\frac{p_x^2}{2m}+\frac{p_y^2}{2m}+V(\bold r)+m_z(|g_\uparrow\rangle\langle g_\uparrow|-|g_\downarrow\rangle\langle g_\downarrow|)\nonumber\\
&&-\bigr\{\bigr[M_x(x)+iM_y(y)\bigr]|g_\uparrow\rangle\langle g_\downarrow|+{\rm H.c.}\bigr\},
\end{eqnarray}
where $V(\bold r)$ is given by Eq.~\eqref{eqn:diople3}, and $M_{x,y}$ are given by Eq.~\eqref{eqn:Raman1}. This is the Eq.~(1) in the main text.

\subsection{Some remarks}

From the above discussions we can see that the atom-light couplings have two effects, namely, inducing the diagonal potentials and driving Raman transitions. The diagonal potentials are approximately spin-independent for $^{40}$K atoms in the parameter regime that the single-photon blue detunings $\Delta_{1,2}$ are much larger than the hyperfine-structure splitting but less than the fine-structure splitting. The two standing wave lasers $\Omega_{1,2}$ generate the conventional square optical lattice, and together with the two plane waves $\tilde\Omega_{1,2}$ the Raman fields are induced. Note that the square lattice and the spatial profile of the Raman fields are determined by the same standing wave lasers. The relative configuration between the Raman fields and square lattice is automatically fixed: the periodicity of the Raman fields is double of the lattice period, and the Raman fields are antisymmetric corresponding to the center of each lattice site [Fig.~\ref{SImodel2} (c)]. This implies that the phase fluctuations in the standing wave lasers, characterized by $\Omega_1=\Omega_0\sin(k_0x+\phi_1^{\rm fluc})$ and $\Omega_2=\Omega_0\sin(k_0y+\phi_2^{\rm fluc})$, only lead to global shift of the lattice and Raman fields as illustrated in Fig.~\ref{SImodel2}(c). The relative spatial profile of $M_{x,y}$ and $V(\bold r)$ is always automatically pinned up and therefore the effective Hamiltonian~\eqref{eqn:H1} is unchanged. This greatly simplifies the experimental setup in the realization. Compared with the recent experiments of generating 1D spin-orbit coupling in cold atoms through a single pair of lasers \cite{SIWang} (and refs.~[10,13,14] in the main text), the present proposal applies one more pair of lasers to drive another independent Raman transition and uses blue-detuned lasers. Note that both pairs of lasers couple to the same two atomic ground states and no additional optical transitions to other atomic ground states are needed. This implies that the realization of the Hamiltonian~\eqref{eqn:H1} can be directly studied based on the existed experimental platforms.

\section{Tight-binding Model}

Now we derive the tight-binding model from the effective Hamiltonian~\eqref{eqn:H1}. We take that fermions occupy the lowest $s$-orbitals $\phi_{s\sigma}$ ($\sigma=\uparrow,\downarrow$), and consider only the nearest-neighbor hopping terms. The tight-binding Hamiltonian is given by
\begin{eqnarray}\label{eqn:tightbinding1}
H_{\rm TI}=-\sum_{<\vec{i},\vec{j}>,\sigma}t_s\hat c_{\vec{i}\sigma}^{\dag}\hat
c_{\vec{j}\sigma}+\sum_{<\vec{i},\vec{j}>}\bigr(t_{\rm so}^{\vec{i}\vec{j}}\hat c_{\vec{i}\uparrow}^\dag\hat c_{\vec{j}\downarrow}+{\rm H.c.}\bigr)+\sum_{\vec{i}}m_z\bigr(\hat n_{\vec{i}\uparrow}-\hat n_{\vec{i}\downarrow}\bigr),
\end{eqnarray}
where $\vec i=(i_x,i_y)$ is the 2D lattice-site index, the Zeeman term $m_z=\hbar\delta/2$, the particle number operators $\hat n_{\vec{i}\sigma}=\hat c_{\vec{i}\sigma}^\dag\hat c_{\vec{i}\sigma}$, and $t_s$ denotes spin-conserved hopping, given by
\begin{eqnarray}\label{eqn:hoping1}
t_s=\int d^2\bold r\phi^{(\vec j)}_{s\sigma}(\bold r)\bigr[\frac{p_x^2+p_y^2}{2m}+V(\bold r)\bigr]\phi^{(\vec j+\vec 1)}_{s\sigma}(\bold r).
\end{eqnarray}
On the other hand, from the even-parity (relative to the lattice-site center) of the local $s$-orbitals $\phi_{s\sigma}$ ($\sigma=\uparrow,\downarrow$), the periodic terms $M_{x,y}$, which are odd relative to each lattice-site center, do not couple the intrasite orbitals, but lead to spin-flip hopping with the coefficient given by
\begin{eqnarray}\label{eqn:hoping2}
t_{\rm so}^{\vec i\vec j}=\int d^2\bold r\phi^{(\vec i)}_{s\uparrow}(\bold r)M_{x,y}(\bold r)\phi^{(\vec j)}_{s\downarrow}(\bold r),
\end{eqnarray}
representing the induced 2D SO interaction. It can be directly verified that the spin-flip hopping terms due to the Raman fields satisfy
\begin{eqnarray}\label{eqn:hoping3}
t_{\rm so}^{j_x,j_x\pm1}=\pm(-1)^{j_x}t_{\rm so}^{(0)}, \ t_{\rm so}^{j_y,j_y\pm1}=\pm i(-1)^{j_y}t_{\rm so}^{(0)},
\end{eqnarray}
where $t_{\rm so}^{(0)}=M_0\int d^2\vec r\phi_{s}(x,y)\sin(k_0x)\phi_{s}(x-a,y)$ with $M_0=\Omega_0^2/\Delta$ and $a$ the lattice constant. We note that the staggered property of the spin-orbit term are a consequence of the relative spatial configuration of the lattice and the Raman fields: the periodicity of the Raman fields is double of the lattice period, and the Raman fields are antisymmetric corresponding to the center of each lattice site [Fig.~\ref{SImodel2} (c)]. As pointed out in the previous section, this relative configuration is automatically fixed, since the spatial profiles of both the square lattice and the Raman fields are determined by the same standing wave lasers $\Omega_{1,2}$. These results are stable against lattice oscillations when the phase fluctuations are present in the standing wave lasers in realistic experiments.


The staggered spin-flip hopping terms in Eq.~\eqref{eqn:hoping3} bring two important effects. First, these terms lead to 2D spin-orbit coupling. Moreover, the staggered property implies that the coupling between spin-up and spin-down states transfers $\pi/a$ momentum, which effective shifts the Brillouin zone by half for the spin-down states (relative to the spin-up states). The latter effect can be absorbed by redefining the spin-down operator $\hat c_{\vec{j}\downarrow}\rightarrow e^{i\pi\vec r_j/a}\hat c_{\vec{j}\downarrow}$. We then recast the Hamiltonian into
\begin{eqnarray}\label{eqn:tightbinding2}
H_{\rm TI}&=&-t_s\sum_{<\bar{i},\vec{j}>}(\hat c_{\vec{i}\uparrow}^{\dag}\hat
c_{\vec{j}\uparrow}-\hat c_{\vec{i}\downarrow}^{\dag}\hat
c_{\vec{j}\downarrow})+\sum_{\vec{i}}m_z(\hat n_{\vec{i}\uparrow}-\hat n_{\vec{i}\downarrow})+\nonumber\\
&&+\bigr[\sum_{j_x}t_{\rm so}^{(0)}(\hat c_{j_x\uparrow}^\dag\hat c_{j_x+1\downarrow}-\hat c_{j_x\uparrow}^\dag\hat c_{j_x-1\downarrow})+{\rm H.c.}\bigr]+\nonumber\\
&&+\bigr[\sum_{j_y}it_{\rm so}^{(0)}(\hat c_{j_y\uparrow}^\dag\hat c_{j_y+1\downarrow}-\hat c_{j_y\uparrow}^\dag\hat c_{j_y-1\downarrow})+{\rm H.c.}\bigr].
\end{eqnarray}
It can be seen that the $\pi/a$-momentum transfer between spin-up and spin-down states effectively reverses the hopping coefficient $t_s\rightarrow-t_s$ for the spin-down states. This interesting effect makes the above single-particle model be already nontrivial: it leads to the quantum anomalous Hall effect as studied in the main text.

\section{Experimental parameters}

Finally we provide numerical estimate for parameters in the realistic experiments. The upper value of the detunings is limited by the fine-structure splitting and can be taken as $2\pi\times1.7$THz for $^{40}$K atoms, which is actually the parameter regime in the recent experiment~\cite{SIWang}. Therefore in this proposal we take the blue detunings to be $\Delta_{1,2}\sim2\pi\times1.7$THz for the $^{40}$K atoms, which gives the recoil energy $E_R/\hbar\sim\hbar k_0^2/2m=2\pi\times8.5$kHz using lasers of wavelength $764$nm to form the square lattice. The typical difference in $\Delta_{1,2}$ is about $\Delta_1-\Delta_2\sim100$MHz \cite{SIRMPcoldatom}. Taking that $\Omega_0=2\pi\times0.27$GHz (also close to the parameter regime in the experiment~\cite{SIWang}), we have the lattice depth and the amplitude of the Raman fields $V_0=M_0=\Omega_0^2/\Delta=5E_R$, the lattice trapping frequency $\omega=2\pi\times34.9$kHz, and the hopping coefficients $t_{\rm so}^{(0)}/\hbar\sim t_s/\hbar\simeq2\pi\times0.52$kHz. With this parameter regime we find that the bulk gap for QAHE is $E_g=2\pi\times2.08$kHz by setting $\delta=4t_s$, corresponding to the temperature $T\approx0.33T_F\approx100$nK for observation. The BKT temperature for the topological superfluid phase with $U=5t_s, \delta=6t_s,$ and $\mu=-3.5t_s$ is about $T_{\rm BKT}=0.086T_F$.

\end{document}